\documentclass[preprint,groupaddress,nofootinbib,superscriptaddress,amsmath,amssymb]{revtex4}

\usepackage{graphicx}
\usepackage{amsmath,color}

\def\slc#1{\setbox0=\hbox{$#1$}           % set a box for #1
    \dimen0=\wd0                                 % and get its size
    \setbox1=\hbox{/} \dimen1=\wd1               % get size of /
    \ifdim\dimen0>\dimen1                        % #1 is bigger
       \rlap{\hbox to \dimen0{\hfil/\hfil}}      % so center / in box
       #1                                        % and print #1
    \else                                        % / is bigger
       \rlap{\hbox to \dimen1{\hfil$#1$\hfil}}   % so center #1
       /                                         % and print /
    \fi}

\begin{document}

\newcommand{\todo}[1]{(\textbf{\color{red}TODO:} #1)}
\newcommand{\dd}[2]{\frac{{\rm d}#1}{{\rm d}#2}}

\preprint{MPP-2009-170}

\title{Neutrinos from Kaluza--Klein dark matter in the Sun}

\author{Mattias Blennow}
\email{blennow@mppmu.mpg.de}
\affiliation{Max-Planck-Institut f\"ur Physik (Werner-Heisenberg-Institut),\\
F\"ohringer Ring 6, 80805 M\"unchen, Germany}

\author{Henrik Melb\'eus}
\email{melbeus@kth.se}

\author{Tommy Ohlsson}
\email{tommy@theophys.kth.se}
\affiliation{Department of Theoretical Physics, School of Engineering Sciences,\\
Royal Institute of Technology (KTH) -- AlbaNova University Center,\\
Roslagstullsbacken 21, 106 91 Stockholm, Sweden}

\allowdisplaybreaks

\newcommand{\ud}{\mathrm{d}}

\begin{abstract}
We investigate indirect neutrino signals from annihilations of Kaluza--Klein dark matter in the Sun. Especially, we examine a five- as well as a six-dimensional model, and allow for the possibility that boundary localized terms could affect the spectrum to give different lightest Kaluza--Klein particles, which could constitute the dark matter. The dark matter candidates that are interesting for the purpose of indirect detection of neutrinos are the first Kaluza--Klein mode of the $U(1)$ gauge boson and the neutral component of the $SU(2)$ gauge bosons. Using the DarkSUSY and WimpSim packages, we calculate muon fluxes at an Earth-based neutrino telescope, such as IceCube. For the five-dimensional model, the results that we obtained agree reasonably well with the results that have previously been presented in the literature, whereas for the six-dimensional model, we find that, at tree-level, the results are the same as for the five-dimensional model. Finally, if the first Kaluza--Klein mode of the $U(1)$ gauge boson constitutes the dark matter, IceCube can constrain the parameter space. However, in the case that the neutral component of the $SU(2)$ gauge bosons is the LKP, the signal is too weak to be observed.
\end{abstract}

\maketitle

\section{Introduction}

Dark matter (DM) is one of the most active research areas in the interdisciplinary context of particle physics, astroparticle physics, and cosmology today. In recent years, experiments and observations have resulted in compelling evidence for the existence of non-baryonic DM. A good fit to experimental data is obtained if it is assumed that the DM is cold, {\it i.e.}, non-relativistic at the epoch of matter-radiation equality, and that there is a cosmological constant accounting for the observed present expansion of the Universe. An analysis combining the five-year WMAP data, baryon acoustic oscillations, and supernova observations indicates that the DM has a density of $\Omega_{\rm DM} h^2 = 0.1131 \pm 0.0034$ (68 \% CL) \cite{Komatsu:2008hk}. To date, all evidence for DM comes from gravitational interactions, and its particle nature remains unknown. In particular, none of the particles in the current Standard Model (SM) of particle physics could constitute the DM\footnote{Although neutrinos only interact weakly, they are too light to be able to make up more than a small fraction of the DM density.}. Hence, the search for a DM candidate is strongly connected to physics beyond the SM.

An important class of DM candidates constitutes Weakly Interacting Massive Particles (WIMPs), which are weakly interacting particles with masses in the range $\mathcal{O}(1 - 1000) \, {\rm GeV}$. The reason for the importance of WIMPs as DM candidates is that if they are assumed to be thermally produced in the early Universe, the measured present relic DM density can be easily obtained without any further strong assumptions on their properties.

Several currently operating or planned experiments aim to detect WIMPs. This could be done either directly by measuring the recoil of atoms from their elastic scattering on WIMPs \cite{Goodman:1984dc} or indirectly by detecting products of pair annihilations or decays of WIMPs, such as gamma rays, neutrinos, and antimatter. 

Through their interactions with nuclei, WIMPs in the solar system scatter elastically in the Sun, thereby losing energy and becoming trapped in the solar gravitational potential. Therefore, the local DM density in the Sun can exceed the average density by several orders of magnitude, and hence, the Sun could be an interesting source of WIMP annihilation products. In this context, only neutrinos are of interest, since they have a sufficiently low interaction strength to be able to escape the dense medium of the Sun. These neutrinos can be searched for in neutrino telescopes on Earth, such as the AMANDA/IceCube \cite{Abbasi:2009uz,Braun:2009fr,Danninger:2009uf}, ANTARES \cite{Lim:2009jy}, and Super-Kamiokande \cite{Desai:2004pq} experiments.

An interesting type of WIMPs is Kaluza--Klein Dark Matter (KKDM) \cite{Servant:2002aq,Cheng:2002ej}, which arises in certain models of extra dimensions. In particular, Universal Extra Dimensions (UED) models \cite{Appelquist:2000nn}, where all the particles of the SM are allowed to propagate in one or more extra dimensions, could include several different candidates. In these models, a remnant of translational invariance in the extra dimensions gives rise to a multiplicatively conserved quantum number known as Kaluza--Klein parity, forcing the total KK number entering any vertex to be even. This symmetry is similar to $R$-parity in supersymmetric models, and it ensures that the lightest Kaluza--Klein particle (LKP) is stable. If neutral, the LKP could be a good DM candidate, analogously to the lightest supersymmetric particle (LSP) \cite{Martin:1997ns}.

In the UED models, the mass spectrum of the KK particles, and in particular, the identity of the LKP, depends on specific assumptions on the model at the cutoff scale, where the theory breaks down. In the most widely studied five-dimensional model, known as the Minimal Universal Extra Dimensions (MUED) model, the LKP has been determined to be the first KK mode of the $U(1)$ gauge boson by calculating radiative corrections to the tree-level masses of the KK excitations \cite{Cheng:2002iz}. However, as has explicitly been shown in Ref.~\cite{Flacke:2008ne}, other assumptions could change this conclusion.

The most stringent constraints to date set on the five-dimensional MUED model of KKDM are given by the IceCube collaboration \cite{Abbasi:2009vg}, using the 22-string configuration. When completed with in total 80 strings, IceCube will have the ability to set even stronger constraints on the model.

In this paper, we calculate muon fluxes induced by neutrinos coming from annihilation of KKDM in the Sun. We study different LKP candidates in a five- as well as a six-dimensional model. We use the DarkSUSY \cite{Gondolo:2004sc} and WimpSim \cite{Blennow:2007tw} packages to properly treat the capture rates, the interactions of the annihilation products in the Sun, and the propagation of neutrinos to the Earth, and to simulate the interactions of neutrinos in the detector medium on Earth.   

Neutrinos from KKDM annihilations in the Sun have previously been studied in Ref.~\cite{Hooper:2002gs}, and recently in Ref.~\cite{Flacke:2009eu}. In comparison to those works, we use a more careful treatment, and we find relatively small differences when comparing the results. To our knowledge, the only study of neutrinos from KKDM annihilations in the Sun in a six-dimensional model has been performed in Ref.~\cite{Dobrescu:2007ec}. In that work, only the MUED model is investigated, and it is concluded that no observable neutrino telescope signal is obtained.

The rest of the paper is organized as follows: In Sec.~\ref{sec:WimpAnn}, we describe the formalism of WIMP annihilations in the Sun, and how this is implemented in DarkSUSY and WimpSim. In Sec.~\ref{sec:Models}, we introduce the extra-dimensional models that we investigate. In Sec.~\ref{sec:Results}, we present our results in the form of muon fluxes in a neutrino telescope on Earth and compare these to the existing results in the literature. Finally, in Sec.~\ref{sec:Summary}, we summarize and discuss our results. In addition, technical details are provided in the appendices.

\section{Neutrinos from WIMP annihilations}\label{sec:WimpAnn}

\subsection{The annihilation rate}

The evolution of the total number of WIMPs in the interior of the Sun, $N(t)$, is governed by the equation
\begin{equation}\label{eq:WIMPEvol}
\frac{\ud N}{\ud t} = C_c - C_a N^2,
\end{equation}
where $C_c$ is the capture rate of WIMPs in the Sun and
\begin{equation}
C_a = \langle \sigma v \rangle \frac{V_2}{V_1^2}.
\end{equation}
Here, $\sigma$ is the total WIMP pair annihilation cross section, $v$ is the relative WIMP velocity, and $V_j \simeq 2.3 \cdot 10^{17} (j m_{\rm DM} / 10 \, {\rm GeV})^{-3/2} \, {\rm m}^3$ are effective volumes \cite{Griest:1986yu,Gould:1987ir}. The solution to Eq.~\eqref{eq:WIMPEvol} is
\begin{equation}
	N(t) = \sqrt{\frac{C_c}{C_a}} \tanh \left( \frac{t}{\tau} \right),
\end{equation}
where $\tau \equiv 1/\sqrt{C_a C_c}$ sets the equilibration time scale for the process of capture and annihilation. The annihilation rate is given by
\begin{equation}
\Gamma_A = \frac{C_a}{2} N^2 = \frac{C_c}{2} \tanh^2 \left( \frac{t}{\tau} \right),
\end{equation}
which is the expression that is used in DarkSUSY. For times $t \gg \tau$, $\Gamma_A \simeq C_c / 2$, {\it i.e.}, the annihilation rate is determined by the capture rate only. In that case, the only important aspect of the WIMP-WIMP annihilation cross sections are the relative branching ratios into different channels.

\subsection{Capture rate}

The capture rate of WIMPs in the Sun is governed by the elastic WIMP-proton scattering cross section. In the non-relativistic limit, which is applicable for scattering of cold DM, the interactions between WIMPs and atoms can be characterized as either spin-dependent or spin-independent. Spin-independent cross sections are proportional to the square of the atomic number $A$, while spin-dependent cross sections are proportional to the square of the angular momentum, $J(J+1)$ \cite{Bertone:2004pz}. 
 
For the WIMPs that we will consider, the spin-dependent cross section is much larger than the spin-independent one. The capture rate of WIMPs in the Sun is then commonly approximated by the expression \cite{Jungman:1995df}
\begin{equation}\label{eq:capture}
	C_c^{\rm SD} \simeq 3.35 \cdot 10^{18} \, {\rm s}^{-1} \left( \frac{\rho}{0.3 \, {\rm GeV} / {\rm cm}^3} \right) \left( \frac{270 \, {\rm km/s}}{\bar v} \right)^3 \left( \frac{\sigma_{\rm WIMP,p}^{\rm SD}}{10^{-6} \, {\rm pb}} \right) \left( \frac{1 \, {\rm TeV}}{m_{\rm WIMP}} \right)^2,
\end{equation}
where $\rho$ is the local DM density, $\bar v \equiv \langle v^2 \rangle^{1/2}$ is the root-mean-square of the DM velocity dispersion, $\sigma_{\rm WIMP,p}^{\rm SD}$ is the spin-dependent elastic WIMP-proton scattering cross section, and $m_{\rm WIMP}$ is the mass of the WIMP.

In DarkSUSY, a more accurate result for the capture rate is obtained by dividing the Sun into shells and numerically integrating over these shells, as well as the velocity distribution of the WIMPs. We use the DarkSUSY default values for the WIMP population, which is a Gaussian velocity distribution with root-mean-square velocity dispersion $\bar v = 270 \, {\rm km}/{\rm s}$, and local density $\rho = 0.3 \, {\rm GeV}/{\rm cm}^3$. For more details on the calculation of the capture rate, see Ref.~\cite{DSmanual}. 

\begin{figure}[htb]
\centering
\includegraphics[width=0.75\textwidth,clip]{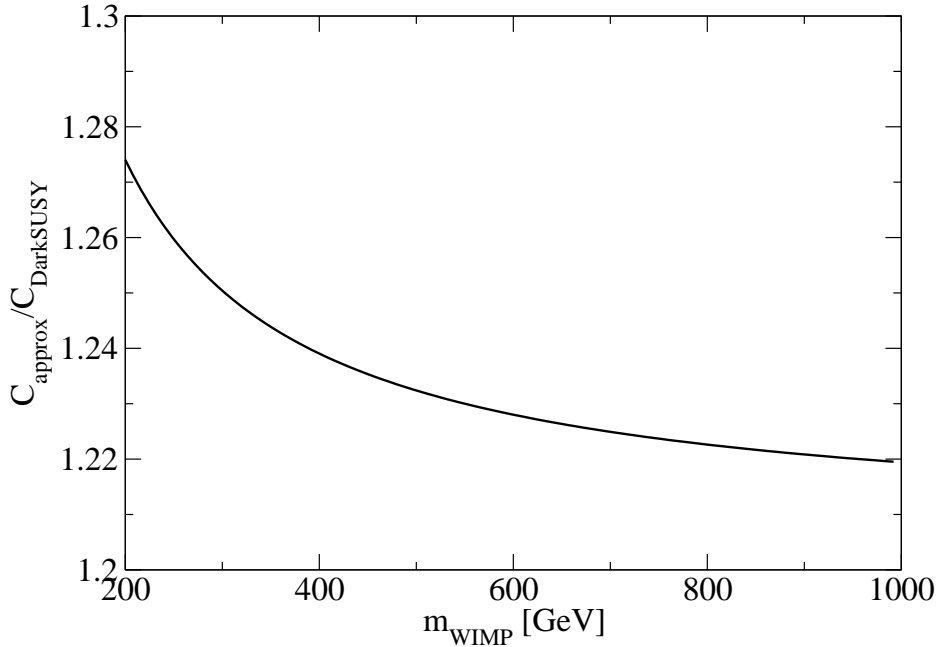}
\caption{The ratio of the capture rate obtained from Eq.~\eqref{eq:capture} to that obtained from DarkSUSY as a function of the WIMP mass $m_{\rm WIMP}$. We have used the spin-dependent elastic scattering cross section $\sigma_{\rm WIMP,p}^{\rm SD} = 1.7 \cdot 10^{-6} \, {\rm pb} \, (1 \, {\rm TeV} / m_{\rm WIMP})^4$.}\label{fig:capture}
\end{figure}

In Fig.~\ref{fig:capture}, we show the ratio of the result obtained from Eq.~\eqref{eq:capture} to the one obtained from DarkSUSY. For this figure, we use $\sigma_{\rm WIMP,p}^{\rm SD} = 1.7 \cdot 10^{-6} \, {\rm pb} \, (1 \, {\rm TeV} / m_{\rm WIMP})^4$, which is the spin-dependent cross section for a $B^{(1)}$ WIMP with $r_{q_R} = 0.1$ (see Eq.~\eqref{eq:ElScBp}). The results from Eq.~\eqref{eq:capture} are larger than the ones from DarkSUSY by about $22 \, \% - 27 \, \%$, with the larger difference at the lowest mass, {\it i.e.}, at the larger cross section.

Since the capture rate is proportional to the WIMP-nucleon scattering cross section, which is also the quantity that is measured in direct detection experiments, the results of neutrino telescope searches and direct detection experiments are expected to be correlated \cite{Halzen:2005ar}. In particular, as stronger limits are set on the WIMP-nucleon cross sections by direct detection experiments, the prospects of detecting neutrinos from WIMP annihilations decrease.

In direct detection experiments, the spin-independent interactions are easier to measure, since a large scattering cross section can be obtained by using heavy target nuclei \cite{Bertone:2004pz}. Hence, the limits on the cross sections are stronger for spin-independent interactions than for spin-dependent ones. The current strongest limits on spin-independent cross sections are set by the CDMS \cite{Ahmed:2008eu} and XENON10 \cite{Angle:2007uj} experiments. The CDMS upper limit on the scattering cross section for a 60 GeV WIMP is $4.6 \cdot 10^{-8}\,$pb at 90 \% C.L. These constraints are strong enough to rule out an observation of neutrinos from annihilations of DM in any running or planned neutrino telescope, if the DM only interacts spin-independently \cite{Halzen:2005ar}.

The most stringent constraints on the spin-dependent cross sections from direct detection experiments are set by XENON10 \cite{Angle:2008we}, COUPP \cite{Behnke:2008zza}, and KIMS \cite{Lee.:2007qn} for scattering on protons and by XENON10 \cite{Angle:2008we}, CDMS \cite{Ahmed:2008eu}, and ZEPLIN III \cite{Lebedenko:2009xe} for scattering on neutrons. The XENON10 results constrain the WIMP-neutron spin-dependent cross section to be less than $5 \cdot 10^{-3} \, {\rm pb}$ for a $30 \, {\rm GeV}$ WIMP, whereas the KIMS results constrain the corresponding cross section for protons to be less than $2 \cdot 10^{-1} \, {\rm pb}$ for an $80 \, {\rm GeV}$ WIMP. In contrast to the limits on the spin-independent cross sections, these constraints are weak enough to allow for a sufficiently strong signal in neutrino telescopes. It is interesting to note that for the case of spin-dependent scattering on protons, IceCube \cite{Abbasi:2009uz} and Super-Kamiokande \cite{Desai:2004pq} have already obtained stronger constraints in some regions of parameter space through indirect detection.

In conclusion, it is especially interesting to use neutrino telescopes to search for DM candidates, which mainly have spin-dependent interactions with nuclei. In particular, scalar DM cannot have spin-dependent interactions, and hence, that class of DM candidates is not interesting in this context.

\subsection{Propagation to the detector}

Once the annihilation rate and branching ratios of the DM annihilations are known, we can proceed by computing the resulting neutrino signals in an Earth based detector. In addition to the geometric scaling of the flux by $1/r^2$, the neutrino flux is subject to interactions with the solar medium as well as matter-enhanced flavor oscillations within the Sun, while the propagation from the solar surface to the Earth is only affected by neutrino oscillations in vacuum.

The consistent theoretical treatment of simultaneous neutrino interactions and oscillations is through the introduction of the neutrino density matrix $\rho(E,r)$, where $E$ is the neutrino energy. This matrix describes the energy-dependent neutrino flavor composition at a distance $r$ from the center of the Sun and follows the evolution equation
\begin{equation}
\dd{\rho}{r} = [H,\rho] - \frac{\rm i}{2} \{\Gamma,\rho\} + \left.\dd{\rho}{r}\right|_{\rm NC} + \left.\dd{\rho}{r}\right|_{\rm CC},
\end{equation}
where $H$ is the neutrino oscillation Hamiltonian, $\Gamma$ is a diagonal matrix containing the interaction rates of the different flavors with the medium, and the two last terms are regeneration terms due to neutral- and charged-current interactions, respectively. In general, the neutral-current regeneration term is simply a decrease of the neutrino energy $E$, since the neutral current is flavor blind, while the charged-current regeneration term projects out the tau-neutrino component of the scattering, since only taus decay before stopping in the Sun. For the specific form of the regeneration terms, see Ref.~\cite{Cirelli:2005gh}.

In our analysis, we use the results of the WimpSim package \cite{Blennow:2007tw}, which provides an event-based Monte Carlo simulation that allows us to compute the neutrino signal in a detector, including all of the effects mentioned above. The oscillations and interactions in the Sun are treated in a consistent framework which can be easily shown to have the same properties as the density matrix evolution described above. Furthermore, the results from WimpSim have been tabulated and are included in DarkSUSY, thus greatly simplifying our implementation. Finally, we compute the resulting muon-antimuon flux at the detector through the standard DarkSUSY routines. This flux can then be compared to the experimental limits in order to rule out models or to see what kind of sensitivity that would be needed to discover the indirect neutrino signal \cite{Wikstrom:2009kw}.

\section{The models}\label{sec:Models}

We investigate the signals of a five- as well as a six-dimensional UED model. For both cases, we choose the models most commonly studied in the literature, though we allow for a more general mass spectrum than that of the MUED models. Each model is based on the SM gauge group $SU(3) \times SU(2) \times U(1)$ and includes the minimal particle content necessary to reproduce the SM in the low-energy limit.

For simplicity, we follow the usual practice of ignoring electroweak symmetry breaking (EWSB) effects. In particular, we treat all the SM particles as massless. We also neglect the Yukawa couplings, which are small compared to the gauge couplings. Since we ignore EWSB effects, we use a gauge where all four components of the Higgs field appear as physical states. Finally, none of the processes that we study involve the self-coupling of the Higgs boson, and hence, we ignore this interaction.

It should be noted that any extra-dimensional field theory is non-renormalizable, and hence, it can only be viewed as an effective theory. Thus, the models that we study are only valid up to some cutoff scale $\Lambda$, at which some more fundamental ultraviolet (UV) completion of the model is needed. The possible effects from physics at the cutoff scale are discussed below. 

\subsection{Five dimensions}

In five dimensions, spinors are four-component objects, as in four dimensions. However, there is no chirality operator \cite{Polchinski:1998rr}, and hence, the Dirac representation is irreducible, {\it i.e.}, all fermions are four-component Dirac fermions. This implies that the simplest choice of geometry for the fifth dimension, the circle $S^1$, does not reproduce the SM in the low-energy limit. This problem can be solved by replacing the circle by the orbifold $S^1 / {\mathbb Z}_2$, which is obtained from $S^1$ by identifying the points $y$ and $-y$, where $y$ is the coordinate along the circle. For the action to be invariant under this transformation, the five-dimensional fields have to be either even or odd in $y$. The KK expansion for an even field is
\begin{equation}
	A^{\rm (even)}(x^\mu,y) = \frac{1}{\sqrt{\pi R}} \left[ A^{(0)} (x^\mu) + \sqrt{2} \sum_{n=1}^\infty \cos \left( \frac{n y}{R} \right) A^{(n)} (x^\mu) \right], 
\end{equation}
and for an odd field
\begin{equation}
	A^{\rm (odd)}(x^\mu,y) = \sqrt{\frac{2}{\pi R}} \sum_{n=1}^\infty \sin \left( \frac{n y}{R} \right) A^{(n)} (x^\mu).
\end{equation}
In particular, an odd field does not have a zero mode, and hence, the field is not present in the low-energy theory, which only contains the zero modes. The left- and right-handed parts of a five-dimensional Dirac field can be given opposite transformation properties, in which case it will have a definite chirality in the low-energy limit. The model can then be made compatible with the SM by introducing such a Dirac field for each chiral fermion in the SM. For a SM fermion $f$, we denote the Dirac fermion corresponding to the left-handed part by $f_D$ and that corresponding to the right-handed part by $f_S$. In general, the weak eigenstates $f_S$ and $f_D$ mix to form massive eigenstates, but the mixing is proportional to the mass of the corresponding SM fermion, and hence, we neglect it.

The orbifolding procedure introduces additional complications to the model. The orbifold $S^1 / {\mathbb Z}_2$ has two fixed points, $y=0$ and $y=\pi R$, where $R$ is the radius of the circle. In general, terms localized to these points, so-called boundary localized terms (BLTs), can be included in the Lagrangian. These terms affect the spectrum and the coupling constants of the model already at tree level. It has been shown that BLTs are generated by radiative corrections \cite{Georgi:2000ks,Cheng:2002iz}, and hence, it is inconsistent not to include them.

The BLTs break translational invariance in the fifth dimension, and hence, the fifth component of the momentum vector is not conserved. However, if the BLTs appear symmetrically between the fixed points, there is a remaining mirror symmetry, and correspondingly, a multiplicatively conserved quantum number, called KK parity \cite{Appelquist:2000nn}. It is defined as $(-1)^n$, where $n$ is the KK number, and it is analogous to $R$-parity in supersymmetric theories \cite{Martin:1997ns}. In particular, it ensures that the LKP is stable. In this work, we always assume that KK parity is conserved.

In contrast to the bulk terms, the BLTs are not determined by the SM parameters. In principle, all such terms that are allowed by the SM gauge symmetry should be included, and {\it a priori}, nothing can be said about these terms, except what constraints can be put on them from experiments \cite{Flacke:2008ne}. Hence, their existence decreases the predictivity of the model. The common solution to this problem is to assume that the BLTs vanish at the cut-off scale $\Lambda$ of the theory. In the resulting model, usually referred to as the MUED model, there are only two free parameters in addition to the SM parameters, namely the compactification radius $R$ and the cut-off scale $\Lambda$. The procedure is similar to the fixing of supersymmetry breaking parameters by making an ansatz at the scale of grand unification, such as mSUGRA \cite{Martin:1997ns}. However, as long as nothing is known regarding the UV completion of the UED model, the MUED ansatz is arbitrary and only made for simplicity.

Our approach to the BLTs is similar to that given in Ref.~\cite{Arrenberg:2008np}. We do not investigate the effects of individual terms, but assume that the spectrum is affected in such a way as to give different LKPs. For simplicity, the possible effects on the couplings are not taken into account.

Gauge fields in higher-dimensional theories have an extra component for each additional dimension. Thus, in the five-dimensional case, each gauge field has a single extra component, $A_5$. From the four-dimensional point of view, the zero mode of this field would appear as a massless scalar, which would be in conflict with experimental results, since no such scalar has been found. Hence, the only phenomenologically viable option is to take $A_5$ to be odd in $y$, so that it does not have a zero mode. On the other hand, the four components $A_\mu$ have to be even in order to reproduce the SM at low energies.

For the Higgs doublet, we use the convention
\begin{equation}\label{eq:Higgs}
	\Phi = \left( \begin{array}{c}
{\rm i} h^+ \\ 
\frac{1}{\sqrt{2}} (h^1 + {\rm i} h^2)
\end{array} \right),
\end{equation}
and the conjugate of $h^+$ is denoted by $h^-$. As in the SM, the zero modes $h^{2(0)}$ and $h^{\pm(0)}$ are eaten by the $Z$ and $W^\pm$ bosons, respectively. For $n>0$, on the other hand, $h^{2(n)}$ and $h^{\pm(n)}$ mix with $Z_5^{(n)}$ and $W_5^{\pm(n)}$, and for each pair, one linear combination is eaten to give mass to the corresponding KK mode of a gauge boson, while the orthogonal combination survives as a physical scalar. However, the mixing is proportional to $v^2 R^2$, where $v$ is the Higgs vacuum expectation value, and since we ignore EWSB effects, there is no mixing. This means that the scalar that is eaten is the fifth component of the gauge boson, while all four components of the Higgs boson appear as physical scalars at each non-zero KK level. 

At each KK level, the electroweak gauge bosons $B^{(n)}$ and $W^{3(n)}$ mix to form two massive eigenstates, $\gamma^{(n)}$ and $Z^{(n)}$, in the same way as in the SM. However, in the UED model, the mass matrix in flavor basis at the $n$th level is given by \cite{Cheng:2002iz}
\begin{equation}
	\left( \begin{array}{cc} \frac{n^2}{R^2} + \delta m_{B^{(n)}}^2 + \frac{1}{4} g_1^2 v^2 & \frac{1}{4} g_1 g_2 v^2 \\ \frac{1}{4} g_1 g_2 v^2 & \frac{n^2}{R^2} + \delta m_{W^{3(n)}}^2 + \frac{1}{4} g_2^2 v^2 \end{array} \right),
\end{equation}
where $\delta m_{B^{(n)}}^2$ and $\delta m_{W^{3(n)}}^2$ are corrections to the tree-level masses. For $n>0$, there are large contributions to the diagonal entries, and the mass matrix is almost diagonal. For $R^{-1} \geq 300 \, {\rm GeV}$, the first-level Weinberg angle is bounded as $\sin \theta_W^{(1)} \lesssim 0.05$, and we can make the approximation $B^{(1)} \simeq \gamma^{(1)}$ and $W^{3(1)} \simeq Z^{(1)}$. We use the basis $(B^{(1)}, W^{3(1)})$, although the $(\gamma^{(1)}, Z^{(1)})$ basis is sometimes used in the literature.

We also ignore the effects of the KK modes above the first level. The relevant parts of the five-dimensional Lagrangian, written in terms of the KK modes, are then \cite{MUEDmanual,Burnell:2005hm}
\begin{eqnarray}\label{eq:L5Dgauge}
	\nonumber \mathcal{L}_{\rm gauge} & = & -\frac{g}{2} f^{abc} F_{\mu\nu}^{(0),a} A^{(1),b\mu} A^{(1),c\nu} \\
\nonumber & & -\frac{g}{2} f^{abc} (\partial_\mu A_\nu^{(1),a} - \partial_\nu A_\mu^{(1),a})(A^{(0),b\mu} A^{(1),c\nu} + A^{(0),c\nu} A^{(1),b\mu}) \\
& & - \frac{g^2}{4} \left[ f^{abc} (A_\mu^{(0),b} A_\nu^{(1),c} + A_\nu^{(0),c} A_\mu^{(1),b}) \right]^2,\\
\label{eq:L5Dfermion}
	\nonumber \mathcal{L}_{\rm fermion} & = & {\rm i} g \left( \overline{f^{(0)}} \gamma^\mu A_\mu^{(1)} P_L f_D^{(1)} + \overline{P_L f_D^{(1)}} \gamma^\mu A_\mu^{(1)} f^{(0)} \right) \\
	& & + {\rm i} g \left( \overline{P_L f_D^{(1)}} \gamma^\mu A_\mu^{(0)} P_L f_D^{(1)} + \overline{P_R f_S^{(1)}} \gamma^\mu A_\mu^{(0)} P_R f_S^{(1)} \right),\\
\label{eq:L5Dscalar}
	\nonumber \mathcal{L}_{\rm scalar} & = & {\rm i} g \left[ A_\mu^{(0)} \left( \Phi^{(1)} \partial^\mu \Phi^{(1)\dagger} - \Phi^{(1)\dagger} \partial^\mu \Phi^{(1)} \right) \right. \\ 
	\nonumber & & \left. + A_\mu^{(1)} \left( \Phi^{(0)} \partial^\mu \Phi^{(1)\dagger} - \Phi^{(1)\dagger} \partial^\mu \Phi^{(0)} + \Phi^{(0)\dagger} \partial^\mu \Phi^{(1)} - \Phi^{(1)} \partial^\mu \Phi^{(0)\dagger} \right) \right] \\
	\nonumber & & + g^2 \left[ A_\mu^{(1)} A^{(0)\mu} \left( \Phi^{(1)} \Phi^{(0)\dagger} + \Phi^{(1)\dagger} \Phi^{(0)} \right) \right. \\
	& & \left. + A_\mu^{(1)} A^{(1)\mu} |\Phi^{(0)}|^2 + A_\mu^{(0)} A^{(0)\mu} |\Phi^{(1)}|^2 \right].
\end{eqnarray}
The symbols $f^{abc}$ are the structure constants for the gauge group, and in Eqs.~\eqref{eq:L5Dfermion} and \eqref{eq:L5Dscalar}, $A \equiv A^a T^a$, where $T^a$ are the generators of the gauge group. For completeness, the corresponding Feynman rules are presented in Appendix \ref{sec:FeynmanRules5D}.

\subsubsection{Dark matter candidates}

Once the boundary terms are known, the spectrum of the model can be worked out, and the identity of the LKP can be resolved. In general, the LKP could be the first KK mode of any particle in the SM, or of the graviton. The particle content of the first KK level is equal to that of the SM, except that the left- and right-handed Weyl fermions are replaced by two Dirac fermions with the corresponding quantum numbers and that all the components of the Higgs doublet are physical. Since the LKP has to be neutral with respect to the electromagnetic as well as the color interactions in order to be a good DM candidate, the list of interesting possibilities reduces to the neutrinos, the two neutral components of the Higgs doublet, and the $B$ and $W^3$ bosons. As mentioned above, scalar DM is not interesting for the purpose of neutrino telescope detection, and hence, we will not consider the Higgs bosons as DM candidates. In the context of the MUED model, the spin-independent cross section for KK neutrinos scattering on protons \cite{Servant:2002hb} is larger than the limits set by direct detection experiments, and hence, this particle is ruled out as a DM candidate. The same conclusion holds in our model, provided that we do not modify the couplings. We also note that although the first KK mode of the graviton could be the LKP, this particle interacts very weakly, and its phenomenology is very different from that of WIMPs \cite{Hooper:2007qk}. Thus, we are left with the first KK modes of the $B$ and $W^3$ gauge bosons.

In the five-dimensional MUED model, the $B^{(1)}$ is the LKP \cite{Cheng:2002iz}. Detailed calculations of the relic density, including the effects of coannihilations, have shown that the $B^{(1)}$ could give rise to the relic density measured by WMAP if its mass is roughly in the range $500 \, {\rm GeV} - 1600 \, {\rm GeV}$ \cite{Burnell:2005hm,Kong:2005hn}. For the five-dimensional model, neutrino signals from the $B^{(1)}$ were considered in Ref.~\cite{Hooper:2002gs} and signals from $B^{(1)}$ and $W^{3(1)}$ were considered in Ref.~\cite{Flacke:2009eu}. We will compare the results of these works to our results in Sec.~\ref{sec:Results}.

\subsection{Six dimensions}

In six dimensions, spinors are eight component objects, and as in four dimensions, there is a chirality operator, {\it i.e.}, an $8 \times 8$ matrix $\Gamma_7$ fulfilling $\Gamma_7^2 = 1$ and $\{ \Gamma_7,\Gamma_M \} = 0$, where $\Gamma_M$ are the six gamma matrices in six dimensions \cite{Polchinski:1998rr}. This operator, which is analogous to $\gamma_5$ in four dimensions, makes it possible to define a six-dimensional chirality. The operators projecting onto the states of definite chirality (which are usually referred to as states of $+$/$-$ chirality) are $P_\pm = (1\pm\Gamma_7)/2$. The states of definite chirality are four-component Dirac spinors. Thus, the fermions can be defined as Dirac spinors, and the situation is then the same as in five dimensions.

The condition of anomaly cancellation in the six-dimensional model places constraints on the chirality assignments for the fermions \cite{Dobrescu:2001ae}. Most notably, this condition forces the number of generations to be an integer multiple of three, motivating the existence of three generations in the SM. It also has implications for the six-dimensional chirality structure of the $SU(2)$ doublets and singlets. The fermion content of the model is \cite{Burdman:2006gy} $Q_+ = (U_+,D_+)$, $U_-$, $D_-$, $L_+ = (N_+,E_+)$, and $E_-$, where the $SU(2)$ doublets have left-handed zero-modes, and the singlets have right-handed zero-modes.

In this work, we study a six-dimensional UED model, where the extra dimensions are compactified on the so-called chiral square, $T^2 / {\mathbb Z}_4$ \cite{Dobrescu:2004zi,Ponton:2005kx,Burdman:2005sr}. This orbifold can be obtained by starting from a square with side length $L$ and identifying the points $(y,0)$ and $(0,y)$ as well as the points $(y,L)$ and $(L,y)$ for $0 \leq y \leq L$. In this model, there are four possible choices for the boundary conditions of a field \cite{Dobrescu:2004zi}. The KK decomposition is given by 
\begin{equation}
	A(x^\mu,x^4,x^5) = \frac{1}{L} \left[ \delta_{n,0} A^{(0,0)} (x^\mu) + \sum_{j \geq 1} \sum_{k \geq 0} f_n^{(j,k)} (x^4,x^5) A^{(j,k)} (x^\mu) \right],
\end{equation}
where
\begin{equation}
	f_n^{(j,k)} (x^4,x^5) = \frac{1}{1+\delta_{j,0}} \left[ e^{-{\rm i} n \pi / 2} \cos \left( \frac{jx^4+kx^5}{R} + \frac{n \pi}{2} \right) \pm \cos \left( \frac{kx^4-jx^5}{R} + \frac{n \pi}{2} \right) \right].
\end{equation}
Here, the index $n = 0,1,2,3$ indicates which boundary conditions are imposed on the field, and we have defined $R = L/\pi$.
The squared mass for a KK mode with indices $(j,k)$ is $m_{j,k}^2 = (j^2+k^2)/R^2$. Note that only the $n=0$ functions have a zero mode and that there is no $(0,1)$ mode. Hence, for each field, there is a single lightest KK mode, with indices $(1,0)$ and mass $1/R$. We call this mode the first KK mode.

As the $S^1 / {\mathbb Z}_2$ orbifold in five dimensions, the chiral square $T^2 / {\mathbb Z}_4$ has fixed points, $(0,0)$, $(L,L)$, and $(0,L)$. These fixed points induce BLTs, and break momentum conservation in the extra dimensions to the conservation of KK parity, which in this model is defined as $(-1)^{j+k}$. In the same way as in the five-dimensional model, the conservation of KK parity ensures the stability of the LKP.

In six dimensions, the gauge fields have two additional components. Similar to the five-dimensional case, one linear combination of these fields is eaten at each KK level to give mass to the corresponding KK mode of the four-dimensional components. However, the other linear combination appears as a physical scalar, known as an adjoint scalar, at each non-zero KK level. The adjoint scalar corresponding to the gauge field $A$ is denoted $A_H$. Note that the index $H$ is not a running index.

The KK-decomposed Lagrangian density has been derived in Ref.~\cite{Burdman:2005sr}. We are only interested in interactions involving zero modes and the first KK modes. Neglecting all other modes simplifies the Lagrangian dramatically. Since $k=0$ in all the modes that are left, we suppress this index. We use a gauge in which the linear combination of the extra components of the gauge fields that are eaten vanishes, and the four-dimensional part is massive. Hence, the fields that are denoted $A_G$ in Ref.~\cite{Burdman:2005sr} vanish. As in the five-dimensional model, we ignore EWSB effects and Yukawa couplings.

The modifications to the couplings from the integration over the extra dimensions are encoded in the $\delta$ symbols, which are defined as
\begin{equation}
\delta_{n_1,\ldots,n_r}^{j_1,\ldots,j_r} \equiv \frac{1}{L^2} \int_0^L \ud x^4 \int_0^L \ud x^5 f_{n_1}^{j_1} \cdots f_{n_r}^{j_r}.
\end{equation}
The order of the indices does not matter, as long as the upper and lower indices are shifted in the same way. Only the $n=0$ functions have zero modes, and as we only consider $j_r \in \{0,1\}$, this means that $n_r > 0$ implies that $j_r=1$. Finally, it holds that $\delta_{n_1,n_2,0,0}^{j_1,j_2,j_3,0} = \delta_{n_1,n_2,0}^{j_1,j_2,j_3}$.

Using the properties summarized above, we obtain the interaction Lagrangian
\begin{eqnarray}
\nonumber \mathcal{L}_{\rm gauge} & = & -g f^{abc} \delta_{0,0,0}^{j_1,j_2,j_3} A_\mu^{(j_1),a} A_\nu^{(j_2),b} \partial^\mu A^{(j_3),c \nu} + \Big( \frac{g}{2} f^{abc} A_H^{(1),a} (\partial^\mu A_H^{(1),b}) A_\mu^{(0),c} + {\rm h.c.} \Big) \\
\nonumber & & - \frac{g^2}{4} f^{abc} f^{ade} \delta_{0,0,0,0}^{j_1,j_2,j_3,j_4} A_\mu^{(j_1),b} A_\nu^{(j_2),c} A^{(j_3),d \mu} A^{(j_4),e \nu}\\
& & - \frac{g^2}{2} f^{abc} f^{ade} A_H^{(1),c} A_H^{(1),e} A_\mu^{(0),b} A^{(0),d \mu},\\
\label{eq:L6DfermionL}\mathcal{L}_{{\rm fermion},+} & = & g \delta_{0,0,0}^{j_1,j_2,j_3} \overline{f_D^{(j_1)}} A_\mu^{(j_2)} \gamma^\mu f_D^{(j_3)} + g \overline{f_S^{(1)}} A_\mu^{(0)} \gamma^\mu f_S^{(1)} + \Big( {\rm i} g \overline{f_D^{(0)}} A_H^{(1)} f_S^{(1)} + {\rm h.c.} \Big),\\
	\label{eq:L6DfermionR}\mathcal{L}_{{\rm fermion},-} & = & g \delta_{0,0,0}^{j_1,j_2,j_3} \overline{f_S^{(j_1)}} A_\mu^{(j_2)} \gamma^\mu f_S^{(j_3)} + g \overline{f_D^{(1)}} A_\mu^{(0)} \gamma^\mu f_D^{(1)} + \Big( {\rm i} g \overline{f_S^{(0)}} A_H^{(1)} f_D^{(1)} + {\rm h.c.} \Big),\\
	\label{eq:L6Dscalar}\nonumber \mathcal{L}_{\rm scalar} & = & \Big( {\rm i} g \delta_{0,0,0}^{j_1,j_2,j_3} \Phi^{(j_1)\dagger} A_\mu^{(j_2)} \partial^\mu \Phi^{(j_3)} + {\rm h.c.} \Big) + g^2 \delta_{0,0,0,0}^{j_1,j_2,j_3,j_4} \Phi^{(j_1)\dagger} A_\mu^{(j_2)} A^{(j_3),\mu} \Phi^{(j_4)} \\
& & - \, g^2 \Phi^{(0)\dagger} A_H^{(1)} A_H^{(1)} \Phi^{(0)},
\end{eqnarray}
where $j_i \in \{0,1\}$, $i=1,2,3,4$, and in Eqs.~\eqref{eq:L6DfermionL}, \eqref{eq:L6DfermionR}, and \eqref{eq:L6Dscalar}, $A \equiv A^a T^a$, where $T^a$ are the generators of the gauge group. The Feynman rules for this model are the same as in the five-dimensional case, with the addition of new interactions involving the adjoint scalars. These new rules are listed in Appendix \ref{sec:FeynmanRules6D}.

\subsubsection{Dark matter candidates}

In comparison to the five-dimensional model, each KK level in the six-dimensional model also includes an adjoint scalar for each gauge boson. In particular, the first-level adjoint scalars $B_H^{(1)}$ and $W_H^{3(1)}$ are neutral, and therefore, they are possible DM candidates. However, being scalars, these candidates are not interesting for our purposes. It has been shown that in the MUED version of the model, the adjoint scalar $B_H^{(1)}$ is the LKP \cite{Ponton:2005kx}. Thus, it is necessary to go beyond the MUED scenario in order to possibly obtain a neutrino telescope signal. We conclude that the interesting LKP candidates in the six-dimensional model are the same as in five dimensions, {\it i.e.}, the $B^{(1)}$ and the $W^{3(1)}$.

\section{Results}\label{sec:Results}

The elastic WIMP-proton scattering cross sections for $B^{(1)}$ and $W^{3(1)}$ LKPs in the five-dimensional model have been calculated in Refs.~\cite{Cheng:2002ej} and \cite{Arrenberg:2008np}, respectively. The results are
\begin{equation}\label{eq:ElScBp}
	\sigma_{B^{(1)},{\rm p}}^{\rm SD} \simeq 1.69 \cdot 10^{-6} \, {\rm pb} \left( \frac{0.1}{r_{q_R}} \right)^2 \left( \frac{1 \, {\rm TeV}}{m_{B^{(1)}}} \right)^4,
\end{equation}
\begin{equation}\label{eq:ElScW3p}
	\sigma_{W^{3(1)},{\rm p}}^{\rm SD} \simeq 0.35 \cdot 10^{-6} \, {\rm pb} \left( \frac{0.1}{r_{q_L}} \right)^2 \left( \frac{1 \, {\rm TeV}}{m_{W^{3(1)}}} \right)^4,
\end{equation}
where $r_{q_{L/R}} \equiv (m_{q_{R/L}^{(1)}}-m_{\rm LKP}) / m_{\rm LKP}$. In Eq.~\eqref{eq:ElScBp}, the contributions from $s$- and $t$-channel exchange of left-handed KK quarks has been neglected. These contributions would increase the cross section by about 3 \%, while introducing an additional free parameter, $r_{q_L}$. On the other hand, in Eq.~\eqref{eq:ElScW3p}, there is no contribution from the right-handed KK quarks, since the $W^3$ boson couples only to left-handed chiral fermions. In the six-dimensional model, there are no tree-level contributions to these cross sections from the adjoint scalars, and hence, Eqs.~\eqref{eq:ElScBp} and \eqref{eq:ElScW3p} are valid also for that model.

The relic DM density was calculated in Ref.~\cite{Arrenberg:2008np} for both $B^{(1)}$ and $W^{3(1)}$ in the five-dimensional model. These calculations include coannihilations in all channels, and were performed for a relative mass splitting $r_{q_{R/L}}$ between the LKP and the KK quarks in the region from 0.01 to 0.5. For the $B^{(1)}$ calculations, the MUED spectrum was used for the rest of the first-level KK particles, and for the $W^{3(1)}$, it was assumed that the $W^{\pm(1)}$ are degenerate with $W^{3(1)}$, that the KK gluons are heavier by 20 \%, and that all other first-level KK particles are heavier by 10 \%. The conclusion is, that in order to obtain the measured relic density, the mass of the $B^{(1)}$ should be roughly in the range $500 \, {\rm GeV} - 1600 \, {\rm GeV}$, depending on the relative mass splitting $r_{q_R}$, and the mass of the $W^{3(1)}$ should be roughly in the range $1800 \, {\rm GeV} - 2700 \, {\rm GeV}$. For the $B^{(1)}$, the preferred mass decreases with increasing mass splitting, while for the $W^{3(1)}$, the opposite is true. Since the interactions in the six-dimensional model are the same as those in the five-dimensional one, the only possible difference in the relic density would come from coannihilations involving adjoint scalars. In this work, we ignore that contribution, and assume that the relevant mass intervals for the respective candidates are approximately the same as for the five-dimensional model.

Using the annihilation cross sections presented in Appendix \ref{sec:Crosssections}, we calculate the branching ratios into different final states, which are presented in Tab.~\ref{tbl:BRs}. The branching ratios into a pair of Higgs bosons are small, and give only a small contribution to the muon-antimuon fluxes. Since this channel would introduce additional model dependence, we choose to neglect it in our calculations.
\begin{table}
\begin{tabular}{|l||l|l|l|l|}
\hline Final state & \multicolumn{2}{c|}{$B^{(1)}$} & \multicolumn{2}{c|}{$W^{3(1)}$} \\ 
\hline $r_q$		& 1.0   & 1.3   & 1.0 & 1.3 \\
\hline \hline $\bar u u$ 	& 0.125 & 0.084 & 0.017 & 0.010 \\ 
$\bar d d$        	& 0.008 & 0.006 & 0.017 & 0.010 \\ 
$\bar \nu \nu$    	& 0.011 & 0.013 & 0.005 & 0.005 \\ 
$l^+ l^-$ 			& 0.183 & 0.223 & 0.005 & 0.005 \\ 
$h h$ 				& 0.004 & 0.005 & 0.002 & 0.002 \\ 
$Z Z$ 				& 0.004 & 0.005 & 0.002 & 0.002 \\ 
$W^+ W^-$ 			& 0.010 & 0.012 & 0.866 & 0.908 \\ 
\hline
\end{tabular}
\caption{The branching ratios for pair annihilation of $B^{(1)}$ and $W^{3(1)}$ LKPs. Except for the LKP and the KK quarks, the masses of the KK particles are assumed to be larger than the LKP mass by 10 \%, except for the $W^{\pm(1)}$, which are assumed to be degenerate with $W^{3(1)}$.}\label{tbl:BRs}
\end{table}

We present our final results as fluxes of muons and anti-muons through an imaginary plane in an Earth-based detector. The detector medium is assumed to be water, with the density $1.0 \, {\rm g}/{\rm cm}^3$. The IceCube collaboration has placed limits on this flux \cite{Abbasi:2009uz} for the two annihilation channels $W^+W^-$ and $\bar b b$. The $W^+W^-$ channel, which has been chosen to represent models resulting in hard neutrino energy spectra, can be used to constrain our models. The reason for this is, that in the case that the $W^{3(1)}$ is the LKP, the branching ratio into $W^+W^-$ is around 90 \%, whereas for the $B^{(1)}$, the majority of the annihilations result in pairs of charged leptons, which also give a hard spectrum. In addition, the limits for the $B^{(1)}$ have been investigated by the IceCube collaboration \cite{Abbasi:2009vg}, and it has been found that they differ from the $W^+W^-$ limits by at most 10 \%.

\begin{figure}[htb]
\centering
\includegraphics[width=0.75\textwidth,clip]{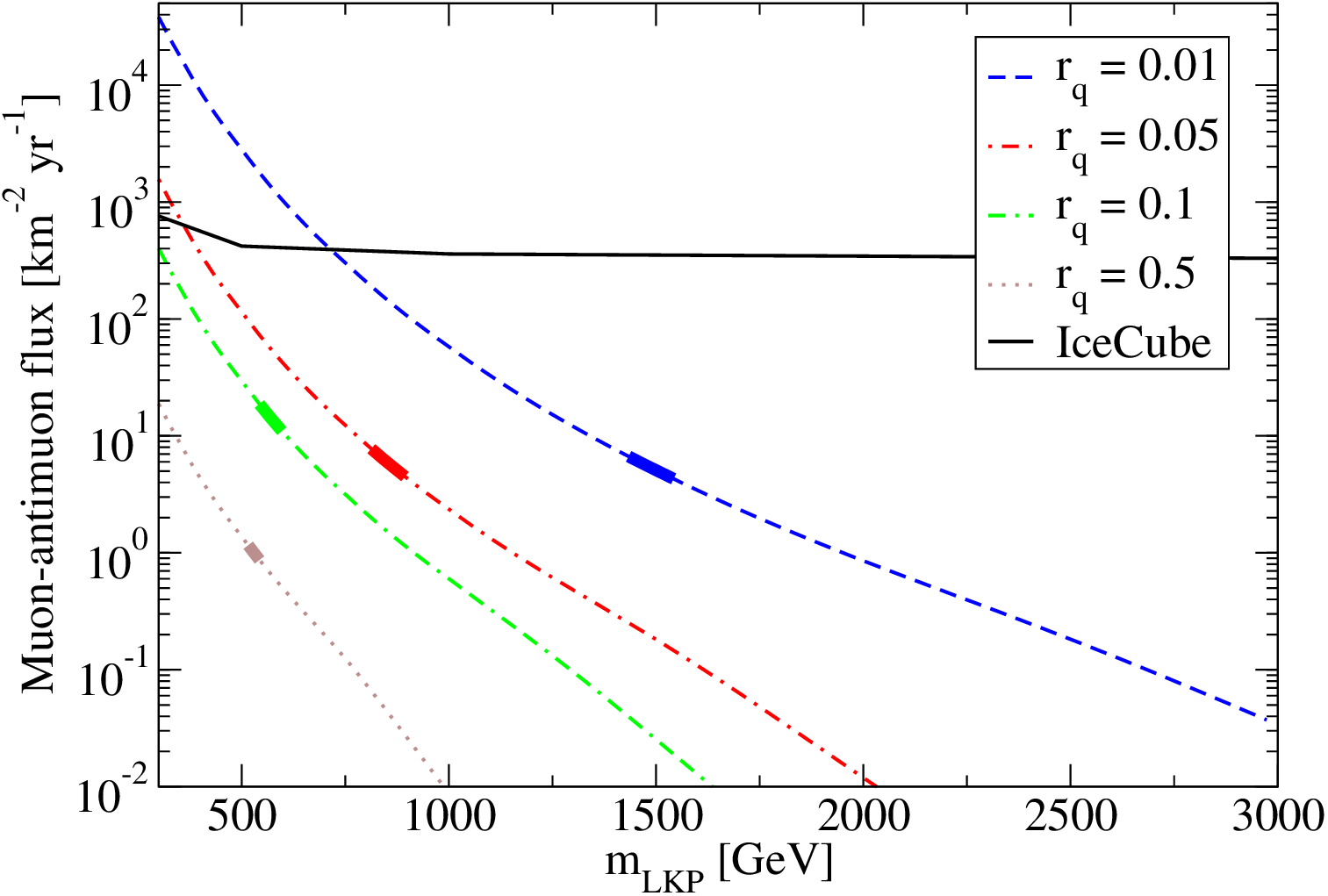}
\caption{The muon-antimuon flux in a detector at Earth as a function of the WIMP mass $m_{\rm WIMP}$ for the case of a $B^{(1)}$ LKP and muon energy threshold $E_\mu^{\rm th} = 1 \, {\rm GeV}$. The dashed (blue), dash-dotted (red), double dash-dotted (green), and dotted (brown) curves represent the relative mass splittings $r_q =$ 0.01, 0.05, 0.1, and 0.5, respectively. Also plotted as the solid (black) curve is the IceCube limit for the hard channel in Ref.~\cite{Abbasi:2009uz}. For each curve, the LKP mass range giving the correct relic abundance is indicated by a thicker segment.}\label{fig:muonfluxBe1}
\end{figure}

\begin{figure}[htb]
\centering
\includegraphics[width=0.75\textwidth,clip]{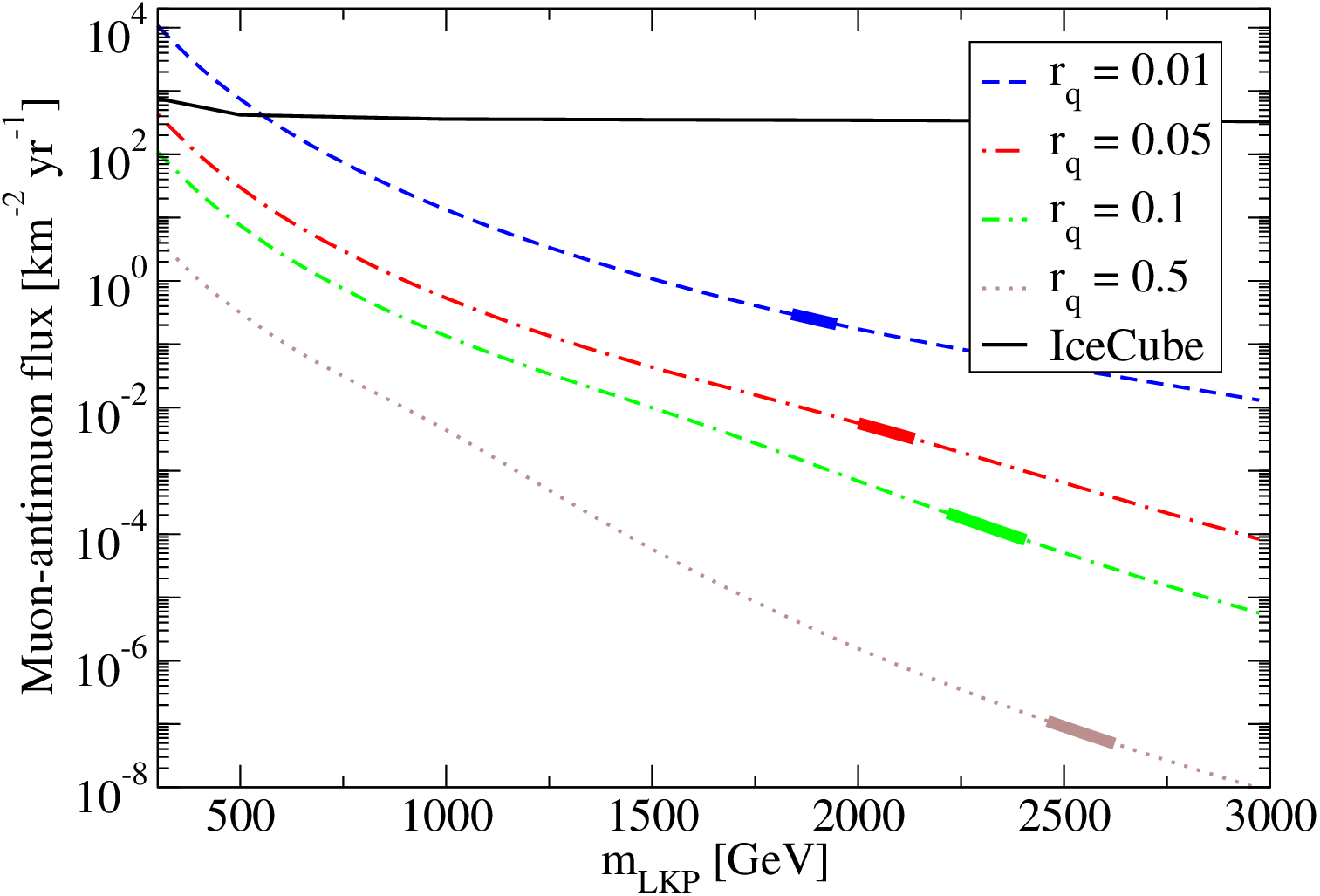}
\caption{The muon-antimuon flux in a detector at Earth as a function of the WIMP mass $m_{\rm WIMP}$ for the case of a $W^{3(1)}$ LKP and muon energy threshold $E_\mu^{\rm th} = 1 \, {\rm GeV}$. The dashed (blue), dash-dotted (red), double dash-dotted (green), and dotted (brown) curves represent the relative mass splittings $r_q =$ 0.01, 0.05, 0.1, and 0.5, respectively. Also plotted as the solid (black) curve is the IceCube limit for the hard channel in Ref.~\cite{Abbasi:2009uz}. For each curve, the LKP mass range giving the correct relic abundance is indicated by a thicker segment.}\label{fig:muonfluxW3e1}
\end{figure}

In Figs.~\ref{fig:muonfluxBe1} and \ref{fig:muonfluxW3e1}, we present the muon-antimuon fluxes for the case that the LKP is the $B^{(1)}$ and the $W^{3(1)}$, respectively. We also include the IceCube limits, and we use the same muon energy threshold as in Ref.~\cite{Abbasi:2009uz}, {\it i.e.}, $E_\mu^{\rm th} = 1 \, {\rm GeV}$. We give the fluxes as functions of the LKP mass, and present results for a number of different values for the relative mass splitting $r_q$ between the LKP and the first-level KK quarks. For all plots, we indicate the mass range for the LKP giving the correct relic density. The masses of the rest of the first-level KK particles only affect the results through the branching ratios, and are less important than the masses of the LKP and the first-level KK quarks. We assume that the $W^{\pm(1)}$ are degenerate with the LKP and that all other particles are heavier than the LKP by 10 \%.

The fluxes for the five- and the six-dimensional models are equal for both LKP candidates. The reason for this is that, for our purposes, the contributions from the adjoint scalars constitute the only difference between the models, and at tree-level, these do not enter the annihilation processes or the scattering of LKPs on nucleons. Hence, each set of fluxes is valid for both of our models.

The fluxes decrease rapidly with increasing LKP mass, an effect which is partly due to the behavior of the scattering cross sections, Eqs.~\eqref{eq:ElScBp} and \eqref{eq:ElScW3p}, which fall off as $m_{\rm LKP}^{-4}$, and partly due to the kinematics of the capture of WIMPs in the Sun, which can be seen from Eq.~\eqref{eq:capture} to be approximately proportional to $m_{\rm LKP}^{-2} \sigma_{\rm WIMP,p}^{\rm sd}$. Taken together, this means that the capture rate falls off approximately as $m_{\rm LKP}^{-6}$. The relative mass splitting $r_q$ mainly affects the fluxes through the scattering cross sections, Eqs.~\eqref{eq:ElScBp} and \eqref{eq:ElScW3p}, which both behave as $r_q^{-2}$. Except for this effect, the mass splitting also affects the branching ratios into different final states, though this is a less important effect.

Comparing the flux from $B^{(1)}$ to that from $W^{3(1)}$, we find that the respective behaviors with mass and KK quark mass splitting are similar, but that the absolute scale of the latter flux is always lower by a factor varying between $1-5$. One of the reasons for this is the lower scattering cross section for the $W^{3(1)}$, which is smaller than that for $B^{(1)}$ by a factor of approximately 5. In addition, the fluxes are affected by the different distribution of final states, which can be seen in Tab.~\ref{tbl:BRs}. While pairs of $B^{(1)}$ particles annihilate mainly into charged leptons, with only a negligible fraction into gauge bosons, the $W^{3(1)}$ annihilates to $W^\pm$ pairs with a very large fraction, and only a small fraction goes to charged leptons. Direct neutrino production has a small but non-zero branching ratio for both LKP candidates. The different annihilation channels will result in two effects. First, the shape of the neutrino spectrum can be very different depending on the channel and a channel with a hard energy spectrum will be easier to distinguish from the background in a neutrino telescope. Thus, for hard spectra, a more stringent bound on the spin-dependent cross section can be put given the same resulting muon flux. The second effect is that of neutrino oscillations, which can increase or decrease the muon flux depending on the annihilation channel. Since oscillations mix the fluxes of different neutrino flavors, a channel giving rise to a muon neutrino dominated spectrum will have its signal decreased by oscillations, and vice versa \cite{Cirelli:2005gh,Blennow:2007tw}.

Our results should be compared to the results for the five-dimensional model that have previously been presented in the literature. In Ref.~\cite{Hooper:2002gs}, the muon-antimuon flux was calculated for the case that the $B^{(1)}$ is the LKP, and in Ref.~\cite{Flacke:2009eu}, these results were updated to include the effects of neutrino oscillations, and the corresponding results for the case that the $W^{3(1)}$ is the LKP were also calculated. In Ref.~\cite{Flacke:2009eu}, the results are presented as rates of muon events per effective detector area for a large volume detector such as IceCube, using a detector depth of 1 km. Due to this depth, the event rate per area is expected to be larger than the flux through a plane in the detector, and therefore, those results are not directly comparable to the results presented in Figs.~\ref{fig:muonfluxBe1} and \ref{fig:muonfluxW3e1}. In order to enable such a comparison, we also calculate the event rates per area using DarkSUSY, and the event rates that we obtained are smaller by a factor of approximately 20 \% - 30 \%. This difference between the results is expected, since we use DarkSUSY to calculate the capture rate, while in Ref.~\cite{Flacke:2009eu}, the approximation given in Eq.~\eqref{eq:capture} is used. As the event rate is proportional to the capture rate, the difference between the event rates is explained by the difference between the capture rates plotted in Fig.~\ref{fig:capture}. It should also be noted that the branching ratios that we obtain for the $W^{3(1)}$ differ from those given in Ref.~\cite{Flacke:2009eu}. We obtain a larger branching ratio into $W^+ W^-$ pairs, while the branching ratios into fermions are smaller. However, we have calculated the muon flux using the branching ratios of Ref.~\cite{Flacke:2009eu}, and have found that the difference compared to our results due to this is not larger than about 10 \%. Finally, the fluxes have been calculated for the five-dimensional MUED model by the IceCube collaboration \cite{Abbasi:2009vg}, and their results are also similar to our results. 

\section{Summary and conclusions}\label{sec:Summary}

In this work, we have studied neutrinos from annihilations of KKDM in the Sun. Using the DarkSUSY and WimpSim packages, we have calculated the flux of neutrino-induced muons and antimuons in an Earth-based neutrino telescope, such as IceCube. We have investigated KKDM in a five-dimensional model, based on the orbifold $S^1/{\mathbb Z}_2$, and in a six-dimensional model, based on the chiral square $T^2/{\mathbb Z}_4$. Rather than restricting ourselves to the MUED models, where the LKP is uniquely determined, we have allowed for the possibility that BLTs might affect the spectrum in such a way as to change the identity of the LKP. The possible DM candidates are then the first KK modes of the neutrinos, the neutral components of the Higgs field, and the $B$ and $W^3$ gauge bosons. In addition, in the six-dimensional model, the adjoint scalars $B_H^{(1)}$ and $W_H^{3(1)}$ are possible candidates. Among these particles, neutrinos are already ruled out by direct detection experiments, while the interactions of scalars with nucleons are too strongly constrained to allow for an observable signature in neutrino telescopes. In five as well as six dimensions, only the $B^{(1)}$ and the $W^{3(1)}$ are left as interesting DM candidates.

For a given LKP mass, the flux is somewhat lower for the case of the $W^{3(1)}$ as the LKP than for the case of the $B^{(1)}$. However, it is important to note that a $B^{(1)}$ LKP is not expected to have the same mass as a $W^{3(1)}$ LKP. Relic density calculations show that the mass of the $B^{(1)}$ should lie in the range from about $500 \, {\rm GeV} - 1600 \, {\rm GeV}$, whereas the $W^{3(1)}$ should be far more massive, in the range from about $1800 \, {\rm GeV} - 2700 \, {\rm GeV}$. Since the capture rate falls off approximately as $m_{\rm LKP}^{-6}$, this implies that the fluxes from the $W^{3(1)}$ are expected to be much smaller. In the relevant range, the flux is predicted to be smaller than $0.2 \, {\rm km}^{-2} \, {\rm yr}^{-1}$. This means that, even assuming a perfect $1 \, {\rm km}^2$ detector, no observable muon fluxes will be generated in IceCube. We wish to emphasize this fact, which has previously not been discussed in the literature. In contrast, in the case that the $B^{(1)}$ is the LKP, IceCube will be able to put constraints on the relevant parts of the parameter space.

Another novel conclusion is that, since the fluxes are equal for the five- and the six-dimensional model for each LKP, it is not possible to distinguish these two models using this indirect detection method.

In conclusion, in the case that the $B^{(1)}$ is the LKP, neutrino telescopes such as IceCube have the ability to set constraints on both the five- and the six-dimensional model that we have studied. On the other hand, if $W^{3(1)}$ is the LKP, the prospects for indirect detection through neutrinos are much worse. Note also, that in this case, the scattering cross section is far below the sensitivity of any currently running or planned direct detection experiments.

\begin{acknowledgments}

We would like to thank Joakim Edsj{\"o} and Matthias Danninger for useful discussions, and Kyoungchul Kong for providing data from relic density calculations.

This work was supported by the European Community through the European Commissions Marie Curie Actions Framework Programme
7 Intra-European Fellowship: Neutrino Evolution [M.B.], the Swedish Research Council (Vetenskapsr{\aa}det),
contract nos.~623-2007-8066 [M.B.], 621-2005-3588, and 621-2008-4210 [T.O.], and the Royal Swedish Academy of Sciences (KVA) [T.O.].

\end{acknowledgments}

\appendix

\section{Feynman rules}

In this appendix, we present the relevant Feynman rules for our models. We do not include QCD interactions or Higgs boson self-interactions, since these do not enter our calculations. The conventions for the components of the Higgs multiplet are given in Eq.~\eqref{eq:Higgs}. We use the hypercharge normalization $Q = T^3 + Y$. The $U(1)$ and $SU(2)$ coupling constants are denoted $g_1$ and $g_2$, respectively.
 All momenta are defined to point inwards. Double lines are used to represent non-zero KK modes, or KK modes with a general index $n$. We only include interactions between zero modes and first level KK modes.

\subsection{Five dimensions}\label{sec:FeynmanRules5D}

\subsubsection*{Gauge boson self-interactions}

\begin{align}
	\nonumber \parbox{50mm}{\includegraphics{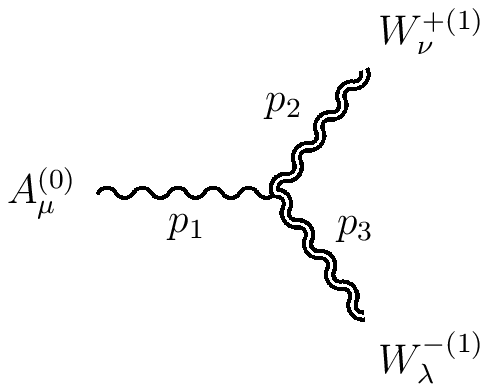}} & = e \left[ g^{\mu\nu} (p_2-p_1)^\sigma + g^{\nu\sigma} (p_3-p_2)^\mu + g^{\sigma\mu} (p_1-p_3)^\nu \right] \\
	\nonumber \parbox{50mm}{\includegraphics{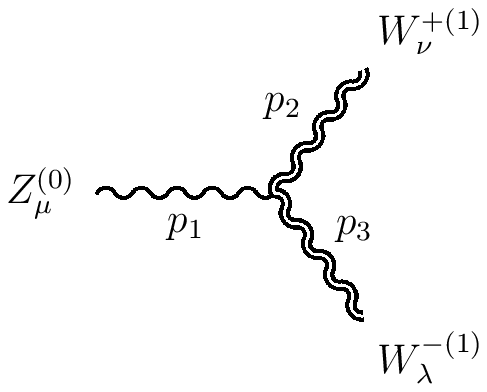}} & = c_w g_2 \left[ g^{\mu\nu} (p_2-p_1)^\sigma + g^{\nu\sigma} (p_3-p_2)^\mu + g^{\sigma\mu} (p_1-p_3)^\nu \right] \\
	\nonumber \parbox{50mm}{\includegraphics{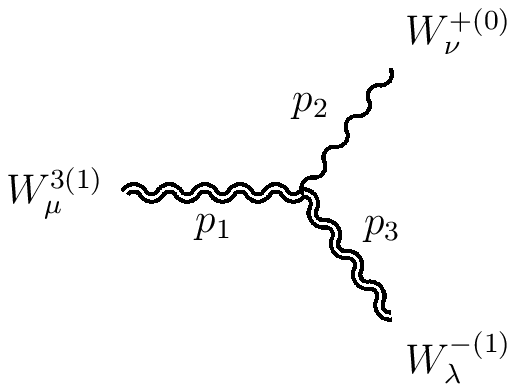}} & = g_2 \left[ g^{\mu\nu} (p_2-p_1)^\sigma + g^{\nu\sigma} (p_3-p_2)^\mu + g^{\sigma\mu} (p_1-p_3)^\nu \right] \\
\label{eq:AAAA} \parbox{50mm}{\includegraphics{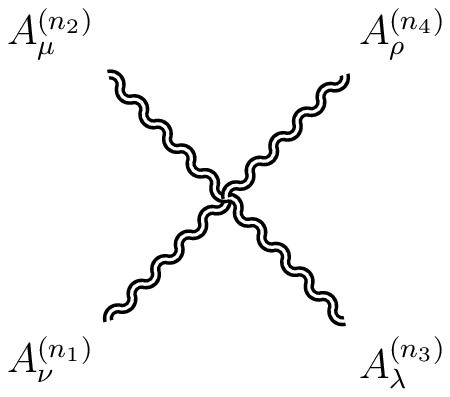}} & = - {\rm i} \mathcal{C}_1 \left(2 g^{\mu\nu} g^{\lambda\rho} - g^{\mu\lambda} g^{\nu\rho} - g^{\mu\rho} g^{\nu\lambda}\right)			
\end{align}
This diagram represents all possible interactions of four gauge bosons, and $A$ stands for any gauge boson. The expressions for the coefficient $\mathcal{C}_1$ for the different vertices are given in Tab.~\ref{tbl:AAAA}.
\begin{table}
\begin{tabular}{|c|c||c|c|}
\hline Vertex & ${\mathcal C}_1$ & Vertex & ${\mathcal C}_1$ \\ 
\hline \hline $A_\mu^{(0)} A_\nu^{(0)} W_\lambda^{+(1)} W_\rho^{-(1)}$ & $e^2$ & $W_\mu^{+(0)} W_\nu^{-(0)} W_\lambda^{3(1)} W_\rho^{3(1)}$ & $g_2^2$ \\ 
\hline $Z_\mu^{(0)} Z_\nu^{(0)} W_\lambda^{+(1)} W_\rho^{-(1)}$ & $c_w^2 g_2^2$ & $A_\mu^{(0)} W_\nu^{3(1)} W_\lambda^{\pm(0)} W_\rho^{\mp(1)}$ & $s_w g_2^2$ \\ 
\hline $A_\mu^{(0)} Z_\nu^{(0)} W_\lambda^{+(1)} W_\rho^{-(1)}$ & $c_w s_w g_2^2$ & $W_\mu^{\pm(0)} W_\nu^{\pm(0)} W_\lambda^{\mp(1)} W_\rho^{\mp(1)}$ & $g_2^2$ \\ 
\hline $Z_\mu^{(0)} W_\nu^{3(1)} W_\lambda^{\pm(0)} W_\rho^{\mp(1)}$ & $c_w g_2^2$ & $W_\mu^{+(0)} W_\nu^{+(1)} W_\lambda^{-(0)} W_\rho^{-(1)}$ & $g_2^2$ \\ 
\hline 
\end{tabular} 
\caption{The expressions for the coefficient ${\mathcal C}_1$ defined in Eq.~\eqref{eq:AAAA} for all possible interactions of four gauge bosons.}\label{tbl:AAAA}
\end{table}

\subsubsection*{Gauge boson-scalar interactions}

These interactions are independent of the KK indices of the interacting modes, except that the vertices have to obey conservation of KK parity. Also, the basis $(A,Z)$ is used for $n=0$, while the basis $(B,W^3)$ is used for $n=1$.

\begin{align}\label{eq:Ahh}
	\parbox{50mm}{\includegraphics{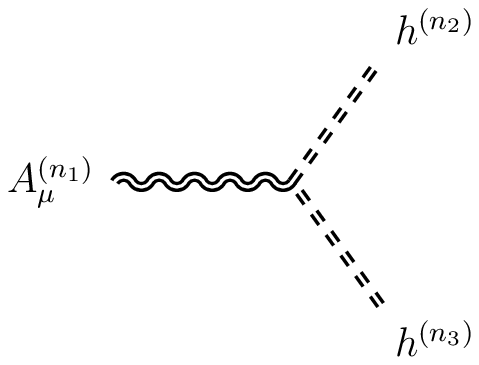}} & = {\rm i} {\mathcal C}_2 P^\mu
\end{align}
This diagram represents all interactions of one gauge boson and two scalars. For all possible interactions of this kind, the expressions for the coefficient ${\mathcal C}_2$ and the momentum $P$ are given in Tab.~\ref{tbl:Ahh}.
\begin{table}
\begin{tabular}{|c|c|c||c|c|c|}
\hline Vertex 			 & ${\mathcal C}_2$ & $P$ & Vertex & ${\mathcal C}_2$ & $P$ \\
\hline \hline $A_\mu h^+ h^-$   & $e$ & $p^+ - p^-$ & $W^3_\mu h^1 h^2$ & $\frac{{\rm i}}{2} g_2$ & $p^2 - p^1$ \\ 
\hline $B_\mu h^+ h^-$   & $\frac{1}{2} g_1$ & $p^+ - p^-$ & $W^-_\mu h^+ h^1$ & $\frac{{\rm i}}{2} g_2$ & $p^+ - p^-$ \\  
\hline $B_\mu h^1 h^2$   & $\frac{{\rm i}}{2} g_1$ & $p^1 - p^2$ & $W^-_\mu h^+ h^2$ & $\frac{1}{2} g_2$ & $p^2 - p^+$ \\  
\hline $Z_\mu h^+ h^-$   & $\frac{1-2s_w^2}{2c_w} g_2$ & $p^+ - p^-$ & $W^+_\mu h^- h^1$ & $\frac{{\rm i}}{2} g_2$ & $p^- - p^1$ \\  
\hline $W^3_\mu h^+ h^-$ & $\frac{1}{2} g_2$ & $p^+ - p^-$ & $W^+_\mu h^- h^2$ & $\frac{1}{2} g_2$ & $p^- - p^2$ \\  
\hline $Z_\mu h^1 h^2$   & $\frac{1}{2c_w} g_2$ & $p^2 - p^1$ & & & \\ 
\hline
\end{tabular} 
\caption{The expressions for the coefficient ${\mathcal C}_2$ and the momentum $P$ defined in Eq.~\eqref{eq:Ahh} for all possible interactions of one gauge boson and two scalars. The interactions are independent of the KK indices of the interacting modes, except that the vertices have to obey conservation of KK parity. The basis $(A,Z)$ is used for $n=0$, while the basis $(B,W^3)$ is used for $n=1$, {\it i.e.}, $A=A^{(0)}$ and $Z=Z^{(0)}$, while $B=B^{(1)}$ and $W^3=W^{3(1)}$.}\label{tbl:Ahh}
\end{table}

\begin{align}\label{eq:AAhh}
	\parbox{50mm}{\includegraphics{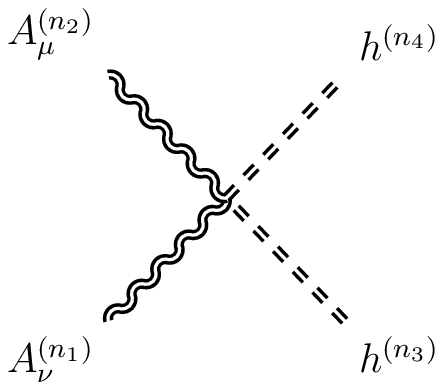}} & = {\rm i} {\mathcal C}_3 g^{\mu\nu}
\end{align}
This diagram represents all interactions involving KK modes of two gauge bosons and two scalars. For all possible interactions of this kind, the expressions for the coefficient ${\mathcal C}_3$ are given in Tab.~\ref{tbl:AAhh}.
\begin{table}
\begin{tabular}{|c|c||c|c||c|c|}
\hline Vertex & ${\mathcal C}_3$ & Vertex & ${\mathcal C}_3$ & Vertex & ${\mathcal C}_3$ \\
\hline \hline $A_\mu A_\nu h^+ h^-$ & $2e^2$ & $W^3_\mu W^3_\nu h^2 h^2$ & $\frac{1}{2} g_2^2$ & $A_\mu W^-_\nu h^+ h^1$ & $\frac{\rm i}{2} e g_2$ \\ 
\hline $A_\mu B_\nu h^+ h^-$ & $e g_1$ & $A_\mu Z_\nu h^+ h^-$ & $\frac{1-2s_w^2}{c_w} e g_2$ & $B_\mu W^-_\nu h^+ h^1$ & $\frac{\rm i}{2} g_1 g_2$ \\ 
\hline $B_\mu B_\nu h^+ h^-$ & $\frac{1}{2} g_1^2$ & $A_\mu W^3_\nu h^+ h^-$ & $e g_2$ & $A_\mu W^-_\nu h^+ h^2$ & $-\frac{1}{2} e g_2$ \\
\hline $B_\mu B_\nu h^1 h^1$ & $\frac{1}{2} g_1^2$ & $B_\mu Z_\nu h^+ h^-$ & $s_w \frac{1-2s_w^2}{c_w^2} g_2^2$ & $B_\mu W^-_\nu h^+ h^2$ & $-\frac{1}{2} g_1 g_2$ \\ 
\hline $B_\mu B_\nu h^2 h^2$ & $\frac{1}{2} g_1^2$ & $B_\mu W^3_\nu h^+ h^-$ & $\frac{1}{2} g_1 g_2$ & $Z_\mu W^+_\nu h^- h^1$ & $\frac{\rm i}{2} e g_1$ \\ 
\hline $Z_\mu Z_\nu h^+ h^-$ & $\frac{(1-2s_w^2)^2}{2 c_w^2} g_2^2$ & $B_\mu Z_\nu h^1 h^1$ & $-\frac{1}{2c_w} g_1 g_2$ & $Z_\mu W^+_\nu h^- h^2$ & $\frac{1}{2} e g_1$ \\ 
\hline $Z_\mu W^3_\nu h^+ h^-$ & $\frac{1-2s_w^2}{2 c_w} g_2^2$ & $B_\mu W^3_\nu h^1 h^1$ & $-\frac{1}{2} g_1 g_2$ & $Z_\mu W^-_\nu h^+ h^1$ & $-\frac{\rm i}{2} e g_1$ \\ 
\hline $W^3_\mu W^3_\nu h^+ h^-$ & $\frac{1}{2} g_2^2$ & $B_\mu Z_\nu h^2 h^2$ & $-\frac{1}{2c_w} g_1 g_2$ & $Z_\mu W^-_\nu h^+ h^2$ & $\frac{1}{2} e g_1$ \\
\hline $Z_\mu Z_\nu h^1 h^1$ & $\frac{1}{2 c_w^2} g_2^2$ & $B_\mu W^3_\nu h^2 h^2$ & $-\frac{1}{2} g_1 g_2$ & $W^+_\mu W^-_\nu h^+ h^-$ & $\frac{1}{2} g_2^2$ \\ 
\hline $Z_\mu W^3_\nu h^1 h^1$ & $\frac{1}{2 c_w} g_2^2$ & $A_\mu W^+_\nu h^- h^1$ & $-\frac{\rm i}{2} e g_2$ & $W^+_\mu W^-_\nu h^1 h^1$ & $\frac{1}{2} g_2^2$ \\ 
\hline $W^3_\mu W^3_\nu h^1 h^1$ & $\frac{1}{2} g_2^2$ & $B_\mu W^+_\nu h^- h^1$ & $-\frac{\rm i}{2} g_1 g_2$ & $W^+_\mu W^-_\nu h^2 h^2$ & $\frac{1}{2} g_2^2$ \\ 
\hline $Z_\mu Z_\nu h^2 h^2$ & $\frac{1}{2 c_w^2} g_2^2$ & $A_\mu W^+_\nu h^- h^2$ & $-\frac{1}{2} e g_2$ & & \\
\hline $Z_\mu W^3_\nu h^2 h^2$ & $\frac{1}{2 c_w} g_2^2$ & $B_\mu W^+_\nu h^- h^2$ & $-\frac{1}{2} g_1 g_2$ & & \\
\hline
\end{tabular} 
\caption{The expressions for the coefficient ${\mathcal C}_3$ defined in Eq.~\eqref{eq:AAhh} for all possible interactions of two gauge bosons and two scalars. The interactions are independent of the KK indices of the interacting modes, except that the vertices have to obey conservation of KK parity. The basis $(A,Z)$ is used for $n=0$, while the basis $(B,W^3)$ is used for $n=1$, {\it i.e.}, $A=A^{(0)}$ and $Z=Z^{(0)}$, while $B=B^{(1)}$ and $W^3=W^{3(1)}$.}\label{tbl:AAhh}
\end{table}

\subsubsection*{Gauge boson-fermion interactions}

\begin{align*}
	\parbox{45mm}{\includegraphics{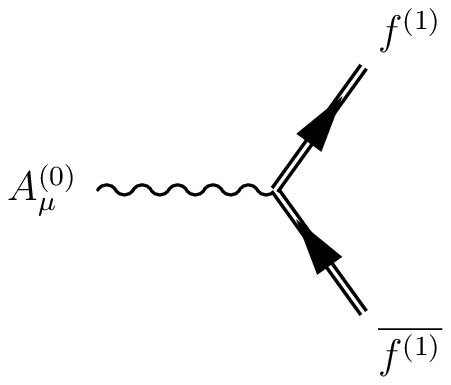}} & = {\rm i} Q_f e \gamma^\mu & \parbox{45mm}{\includegraphics{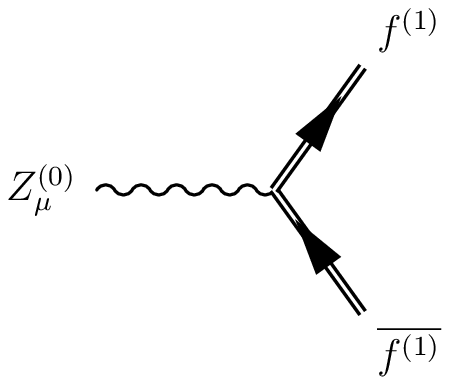}} & = {\rm i} \frac{g_2}{c_w} \left(c_w^2 T_f^3 - s_w^2 Y_f\right) \gamma^\mu \\	
	\parbox{45mm}{\includegraphics{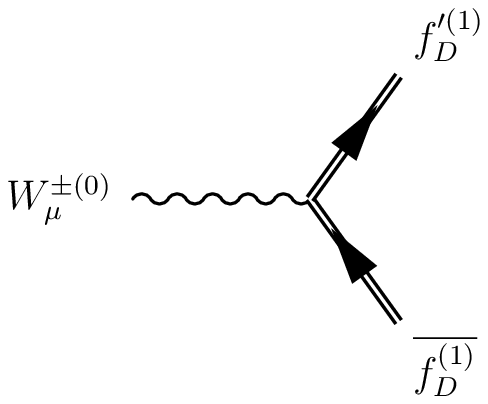}} & = {\rm i} \frac{g_2}{\sqrt{2}} \gamma^\mu & \parbox{45mm}{\includegraphics{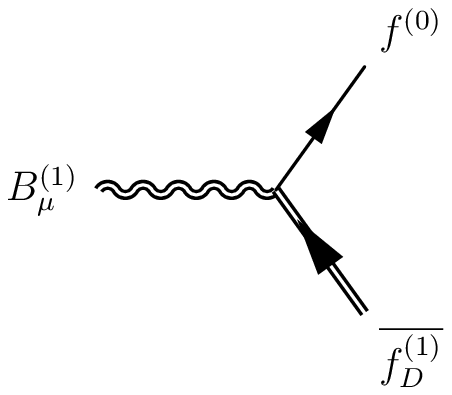}} & = {\rm i} g_1 Y_f \gamma^\mu P_L
\end{align*}

\begin{align*}
	\parbox{50mm}{\includegraphics{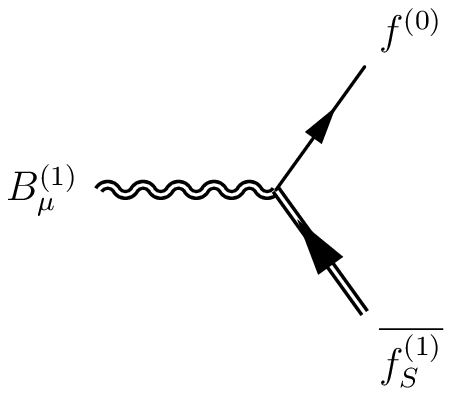}} & = {\rm i} g_1 Y_f \gamma^\mu P_R & \parbox{50mm}{\includegraphics{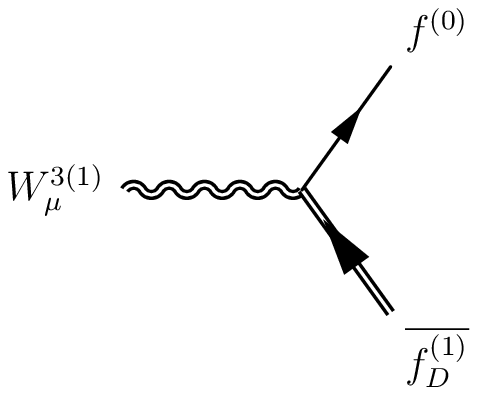}} & = {\rm i} g_2 T_f^3 \gamma^\mu P_L \\	
	\parbox{50mm}{\includegraphics{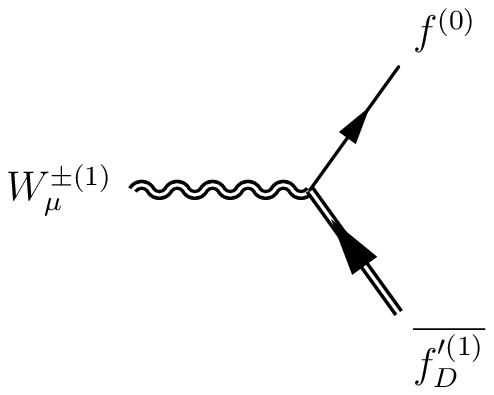}} & = {\rm i} \frac{g_2}{\sqrt{2}} \gamma^\mu P_L	
\end{align*}

\subsection{Six dimensions}\label{sec:FeynmanRules6D}

The only difference between the interactions in the five- and the six-dimensional model is that the six-dimensional model includes additional interactions due to adjoint scalars. Hence, we only present the Feynman rules for these additional interactions in this section.

\subsubsection*{Gauge boson-adjoint scalar interactions}

\begin{align*}
	\parbox{50mm}{\includegraphics{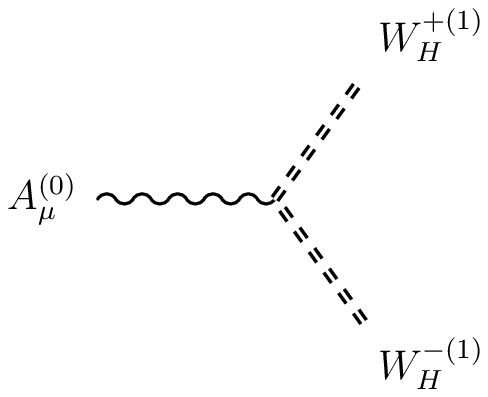}} & = e (p_+ - p_-)^\mu & \parbox{50mm}{\includegraphics{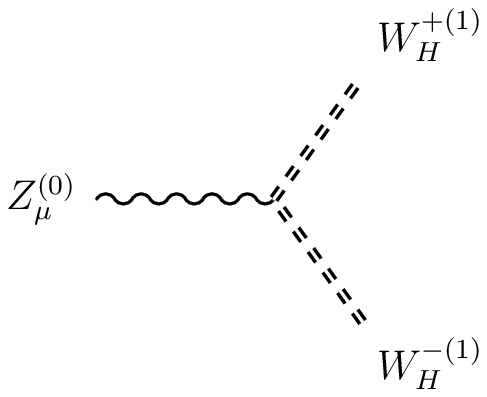}} & = c_w e (p_+ - p_-)^\mu \\
	\parbox{50mm}{\includegraphics{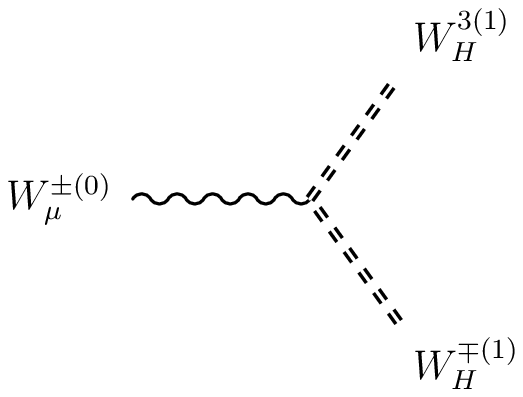}} & = \pm g_2 (p_\mp - p_3)^\mu
\end{align*}

\begin{align}\label{eq:AAAHAH}
	\parbox{50mm}{\includegraphics{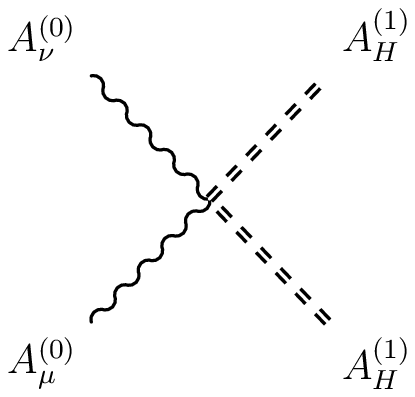}} & = \mathcal{C}_4 g^{\mu\nu}
\end{align}
This diagram represents all possible interactions of two gauge bosons with two adjoint scalars, and $A$ stands for any gauge boson. The expressions for the coefficient $\mathcal{C}_4$ for the different vertices are given in Tab.~\ref{tbl:AAAHAH}.
\begin{table}
\begin{tabular}{|c|c||c|c|}
\hline Vertex & ${\mathcal C}_4$ & Vertex & ${\mathcal C}_4$ \\ 
\hline \hline $A_\mu A_\nu W_H^+ W_H^-$ & $2 e^2$ & $A_\mu W_\nu^\pm W_H^\mp W_H^3$ 	& $-c_w g_2^2$ \\ 
\hline $A_\mu Z_\nu W_H^+ W_H^-$ 	& $2 c_w e g_2$ & $W_\mu^+ W_\mu^- W_H^3 W_H^3$ 		& $2 g_2^2$ \\ 
\hline $A_\mu W_\nu^\pm W_H^\mp W_H^3$ 	& $-e g_2$ & $W_\mu^\pm W_\nu^\pm W_H^\mp W_H^\mp$ 	& $-2 g_2^2$ \\ 
\hline $Z_\mu Z_\nu W_H^+ W_H^-$ 	& $2 c_w^2 g_2^2$ & $W_\mu^+ W_\nu^- W_H^+ W_H^-$ 		& $g_2^2$ \\ 
\hline 
\end{tabular} 
\caption{The expressions for the coefficient ${\mathcal C}_4$ defined in Eq.~\eqref{eq:AAAHAH} for all possible interactions of two gauge bosons with two adjoint scalars.}\label{tbl:AAAHAH}
\end{table}

\subsubsection*{Scalar-adjoint scalar interactions}

\begin{align}\label{eq:AHAHhh}
	\parbox{50mm}{\includegraphics{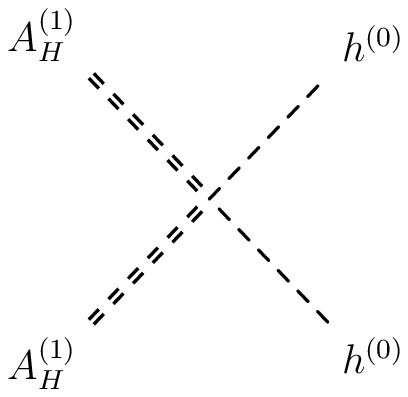}} & = {\rm i} {\mathcal C}_5
\end{align}
This diagram represents all possible interactions of two adjoint scalars with two scalars, and here $A_H$ stands for any adjoint scalar and $h$ stands for any component of the Higgs field. The expressions for the coefficient $\mathcal{C}_5$ for the different vertices are given in Tab.~\ref{tbl:AHAHhh}.
\begin{table}
\begin{tabular}{|c|c||c|c|}
\hline Vertex & ${\mathcal C}_5$ & Vertex & ${\mathcal C}_5$ \\ 
\hline \hline $B_H B_H h^+ h^-$ & $-\frac{1}{2} g_1^2$ 		& $B_H W_H^3 h^1 h^1$ 	& $\frac{1}{2} g_1 g_2$ \\ 
\hline $B_H B_H h^1 h^1$ 		& $-\frac{1}{2} g_1^2$ 		& $B_H W_H^3 h^2 h^2$ 	& $\frac{1}{2} g_1 g_2$ \\ 
\hline $B_H B_H h^2 h^2$ 		& $-\frac{1}{2} g_1^2$ 		& $W_H^+ W_H^- h^+ h^-$ & $-\frac{1}{2} g_2^2$ 	\\ 
\hline $B_H W_H^+ h^- h^1$ 		& $\frac{\rm i}{2} g_1 g_2$ 	& $W_H^+ W_H^- h^1 h^1$ & $-\frac{1}{2} g_2^2$ 	\\ 
\hline $B_H W_H^+ h^- h^2$ 		& $\frac{1}{2} g_1 g_2$ 	& $W_H^+ W_H^- h^2 h^2$ & $-\frac{1}{2} g_2^2$ 	\\ 
\hline $B_H W_H^- h^+ h^1$ 		& $-\frac{\rm i}{2} g_1 g_2$ 	& $W_H^3 W_H^3 h^+ h^-$ & $-\frac{1}{2} g_2^2$ 	\\ 
\hline $B_H W_H^- h^+ h^2$ 		& $\frac{1}{2} g_1 g_2$ 	& $W_H^3 W_H^3 h^1 h^1$ & $-\frac{1}{2} g_2^2$ 	\\ 
\hline $B_H W_H^3 h^+ h^-$ 		& $-\frac{1}{2} g_1 g_2$ 	& $W_H^3 W_H^3 h^2 h^2$ & $-\frac{1}{2} g_2^2$ 	\\ 
\hline 
\end{tabular} 
\caption{The expressions for the coefficient ${\mathcal C}_5$ defined in Eq.~\eqref{eq:AHAHhh} for all possible interactions of two adjoint scalars with two scalars.}\label{tbl:AHAHhh}
\end{table}

\subsubsection*{Adjoint scalar-fermion interactions}

\begin{align*}
	\parbox{50mm}{\includegraphics{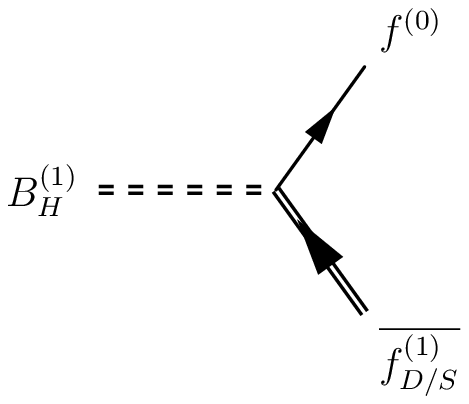}} & = -{\rm i} g_1 Y_f P_{L/R} & \parbox{50mm}{\includegraphics{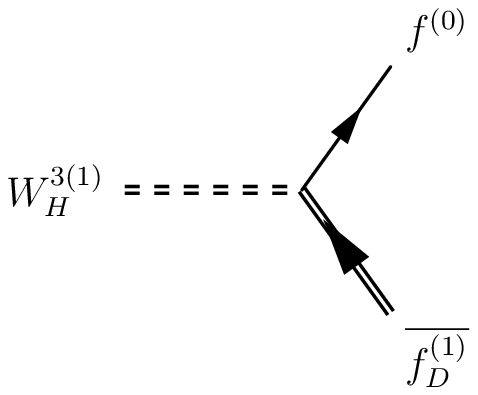}} & = -\frac{{\rm i}g_2}{2} P_L \\
	\parbox{50mm}{\includegraphics{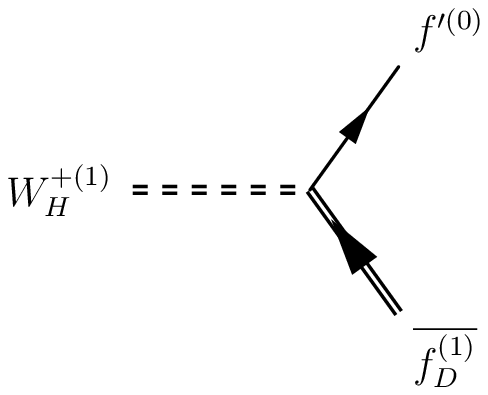}} & = -\frac{{\rm i}g_2}{\sqrt{2}} P_L
\end{align*}

\section{Annihilation cross sections}\label{sec:Crosssections}

In this appendix, we present the relevant WIMP-WIMP annihilation cross sections for the models that we consider. The relevant quantity for our calculations is the cross section times relative velocity, $\sigma v$, in the static limit $s \to 4 m_{\rm WIMP}^2$. We refer to this quantity as the cross section. We let the masses of all first-level KK particles be free parameters. Since we ignore EWSB effects, all the Higgs components appear as physical particles and the symbols $W^\pm$ and $Z$ refer only to the transversal components of those gauge bosons. To calculate the scattering rate into a pair of physical (massive) gauge bosons, the two contributions should be added. For example, the cross section for the annihilation of two $W^{3(1)}$ bosons into a $W^+ W^-$ pair is given by the sum $\sigma (W^{3(1)} W^{3(1)} \to h^+ h^-)+\sigma (W^{3(1)} W^{3(1)} \to W^+ W^-)$.

Since we ignore all SM masses, we simplify the notation by suppressing the KK-indices, {\it i.e.}, $m_{B^{(1)}} \equiv m_B$, {\it etc}. In order to further simplify the notation, we also introduce the functions
\begin{equation}
	f_1(x) = \frac{1}{(1+x)^2},
\end{equation}
\begin{equation}
	f_2(x) = \frac{11-2x+3x^2}{(1+x)^2},
\end{equation}
and
\begin{equation}
	f_3(x) = \frac{1+30x+35x^2+4x^3+6x^4}{x^2(1+x)^2}.
\end{equation}

The annihilation cross sections for the $B^{(1)}$ are
\begin{equation}
	\sigma v (B^{(1)} B^{(1)} \to \bar f f) = \frac{2 N_c Y_f^4 g_1^4}{9\pi m_B^2} f_1 \left(m_{f}^2/m_{B}^2\right),
\end{equation}
where $N_c$ is the number of colors of the fermion $f$,
\begin{eqnarray}
	\nonumber &  & \sigma v (B^{(1)} B^{(1)} \to h^1 h^1) = \sigma v (B^{(1)} B^{(1)} \to h^2 h^2) \\
	\nonumber & = & \frac{1}{2} \sigma v (B^{(1)} B^{(1)} \to h^+ h^-) = \frac{g_1^4}{2304\pi m_B^2} f_2 \left( m_h^2 / m_B^2 \right).
\end{eqnarray}
In the case of degenerate first-level KK masses, the total annihilation cross section is
\begin{equation}
	\sigma v (B^{(1)} B^{(1)})_{\rm tot} = \frac{g_1^4}{\pi m_B^2} \left( \frac{1}{18} \sum_f N_c^f Y_f^4 + \frac{1}{576} \right) \simeq 0.60 \, {\rm pb} \left( \frac{1 \, {\rm TeV}}{m_B} \right)^2.
\end{equation}

For the $W^{3(1)}$, the corresponding results are
\begin{equation}
	\sigma v (W^{3(1)} W^{3(1)} \to \bar f f) = \frac{N_c g_2^4}{72\pi m_{W^3}^2} f_1 \left( m_f^2/m_{W^3}^2 \right),
\end{equation}
\begin{eqnarray}
	\nonumber & & \sigma v (W^{3(1)} W^{3(1)} \to h^1 h^1) = \sigma v (W^{3(1)} W^{3(1)} \to h^2 h^2) \\
	 & = & \frac{1}{2} \sigma v (W^{3(1)} W^{3(1)} \to h^+ h^-) = \frac{g_2^4}{2304\pi m_{W^3}^2} f_2 \left( m_h^2 / m_{W^3}^2 \right),
\end{eqnarray}
\begin{equation}
	\sigma v (W^{3(1)} W^{3(1)} \to W^+ W^-) = \frac{g_2^4}{36\pi m_{W^3}^2} f_3 \left( m_{W^\pm}^2/m_{W^3}^2 \right).
\end{equation}
In the case of degenerate first-level KK masses, the total annihilation cross section is
\begin{equation}
	\sigma v (W^{3(1)} W^{3(1)})_{\rm tot} = \frac{355}{576} \frac{g_2^4}{\pi m_{W^3}^2} \simeq 14 \, {\rm pb} \left( \frac{1 \, {\rm TeV}}{m_B} \right)^2.
\end{equation}

\end{document}